# Thermodynamics of infinite U Hubbard model

R. Kishore [a,*], A. K. Mishra [b]

[a] Instituto Nacional de Pesquisas Espaciais, CP 515, S. J. Campos, SP, 12245-970, Brazil

[b] Institute of Mathematical Sciences, CIT Campus, Chennai, 600113, India

___

**Abstract**

The infinite U Hubbard model, with exclusion of double occupancy of sites, can be considered as a free orthofermion Hamiltonian which is exactly soluble. It is found that the orthofermion distribution function is similar to the mean number of trapped electrons in an impurity in a semiconductor where the double occupancy of the impurity is forbidden and similar to the distribution function of the usual fermions with the chemical potential $\mu$ replaced by $\overline{\mu} = \mu + \ln 2/\beta$. In one dimension, the thermodynamics of free orthofermions gives the known exact results of the infinite U Hubbard model. Thus it shows that at least in one dimension the fermions with exclusion of double occupancy of sites behave as free orthofermions. Since free orthofermions Hamiltonian is exactly soluble in any dimension, it can be employed to ascertain the accuracy of the approximate solutions of the Hubbard model, frequently used for the strongly correlated electron systems like high temperature superconductors.

Keywords: Hubbard model, Strongly correlated electron systems, Orthofermions

___

[*]Corresponding author. Fax: +55-12-3945-6717; e-mail: kishore@las.inpe.br

The Hubbard model is the simplest model to study the strongly correlated electron systems. It is usually written as

$$H = -t \sum_{<ij>\sigma} c_{i\sigma}^{\dagger} c_{j\sigma} + U \sum_{i} n_{i\sigma} n_{i-\sigma} \qquad (1)$$

where $c_{i\sigma}^{\dagger}$ ($c_{i\sigma}$) are creation (annihilation) operators for an electron of spin $\sigma$ ($\pm 1$) on a site i, $n_{i\sigma} = c_{i\sigma}^{\dagger} c_{i\sigma}$ is the electron number operator, t is the nearest neighbor hopping integral, $U$ is the intra-site Coulomb repulsion, and $<ij>$ denotes that the sum is restricted to the nearest neighbors. In case of large U with respect to t, there are formidable problems for its solutions in dimensions higher than one. Even in one dimension, although we have a fairly complete description of the ground state properties, there exist very few exact results at finite temperatures. Exact analytical expressions for the thermodynamical quantities for $U = \infty$ have been derived by many authors [1-3]. However, their methods cannot be extended to higher dimensions.

Recently Mishra [4] has shown that the electrons in the infinite U Hubbard model, with exclusion of double occupancy of sites, can be considered as free orthofermions, obeying the commutation relations

$$c_{i\alpha} c_{j\beta} + c_{j\alpha} c_{i\beta} = 0 \qquad (2)$$

$$c_{i\alpha} c_{j\beta}^{\dagger} = \delta_{\alpha\beta} (\delta_{ij} - \sum_{\sigma} c_{j\sigma}^{\dagger} c_{i\sigma}) \qquad (3)$$

with the Hamiltonian

$$H = -t \sum_{<ij>\sigma} c_{i\sigma}^{\dagger} c_{j\sigma} \qquad (4)$$

Here $\alpha$ and $\beta$ are the spin indices. It should be noted that the commutation relation of Eq. (2) implies that the wave function for the orthofermions is antisymmetric only when the spatial indices i and j are exchanged, keeping unchanged the order of the spin indices. This feature leads to the spin charge decoupling, found in the exact solutions of the Hubbard model in one dimension [5,6].

Using the Fourier transforms in $k$- space

$$c_{k\sigma} = \frac{1}{\sqrt{N}} \sum_i e^{i\mathbf{k}.\mathbf{R}_i} c_{i\sigma} \quad \text{and} \quad \varepsilon_k = -t \sum_{\delta i} e^{i\mathbf{k}.\delta} \tag{5}$$

where $\delta$ is the nearest neighbor vector, and $N$ is total number of sites, the single particle Green´s function [7]

$$G_{k\sigma}(\omega) = \int_{-\infty}^{\infty} dt\, e^{-i\omega t} i\vartheta(t) <[c_{k\sigma}, c_{k\sigma}^{\dagger}(t)]_+> \tag{6}$$

for the free orthofermion Hamiltonian (4) can be obtained exactly. It is given as

$$G_{k\sigma}(\omega) = \frac{1- <c_{k-\sigma}^{\dagger} c_{k-\sigma}>}{\omega - \varepsilon_k} \tag{7}$$

From this Green´s function, one obtains the correlation function [8]

$$<c_{k\sigma}^{\dagger} c_{k\sigma}> = \frac{1- <c_{k-\sigma}^{\dagger} c_{k-\sigma}>}{1+ e^{\beta(\varepsilon_k - \mu)}} \tag{8}$$

It gives the distribution function for orthofermions of wave vector $k$ as

$$n_k = \sum_{\sigma} <c_{k\sigma}^{\dagger} c_{k\sigma}> = \frac{2}{2+ e^{\beta(\varepsilon_k - \mu)}} \tag{9}$$

It should also be noted that the distribution function is similar to the mean number of trapped electrons in an impurity in a semiconductor where the double occupancy of the impurity is forbidden. By rewriting the orthofermion distribution function (9) as $n_k = (1+ e^{\beta(\varepsilon_k - \bar{\mu})})^{-1}$, where $\bar{\mu} = \mu + \ln 2/\beta$, one notices that orthofermions behave as fermions with chemical potential $\mu$ replaced by $\bar{\mu}$. This fact was noticed earlier by Klein [1] in his calculation of the partition function for the infinite $U$ Hubbard model in one dimension.

To study the thermodynamics, we need to know the chemical potential $\mu$ which can

be obtained from

$$n = \sum_{\mathbf{k}} n_{\mathbf{k}} = \sum_{\mathbf{k}} \frac{1}{1 + e^{\beta(\varepsilon_{\mathbf{k}} - \bar{\mu})}} \qquad (10)$$

where n is the number of orthofermions per site. From Eqs. (4) and (7), the internal energy $E = <H>$ is given as

$$E = \sum_{\mathbf{k}} \varepsilon_{\mathbf{k}} n_{\mathbf{k}} = \sum_{\mathbf{k}} \varepsilon_{\mathbf{k}} \frac{1}{1 + e^{\beta(\varepsilon_{\mathbf{k}} - \bar{\mu})}} \qquad (11)$$

The above Eqs. (10) and (11) can determine all the thermodynamical quantities. In one dimension, by changing the summation over $k$ and taking $\varepsilon_k = -2t \cos ka$, where a is the inter-site separation, they can be rewritten as,

$$n = \frac{1}{\pi} \int_0^{\pi} \left[ 1 + \exp(-\beta(2t \cos x + \bar{\mu})) \right]^{-1} dx \qquad (12)$$

$$E = -\frac{2t}{\pi} \int_0^{\pi} \cos x \left[ 1 + \exp(-\beta(2t \cos x + \bar{\mu})) \right]^{-1} dx \qquad (13)$$

The above analytical expressions are exactly equal to those obtained earlier [1-3]. In the zero temperature limit, the self consistent solutions of Eqs. (12) and (13) give the ground state energy, $E_G$, as

$$E_G = -\frac{2t}{\pi} \sin n\pi \qquad (14)$$

which is equal to the exact expression, obtained by Shiba [9] for the infinite $U$ Hubbard model. Thus in one dimension, the fermions with exclusion of double occupancy of sites can be considered as free orthofermions. However, since the free orthofermions Hamiltonian is exactly soluble in any dimension, it can be employed to ascertain the accuracy of the approximate solutions of the Hubbard model, frequently used for the strongly correlated electron systems like high temperature superconductors.

The problem of strongly correlated electron systems has also been studied by using the Gutzwiller variational wave function. However, for the infinite $U$ Hubbard model, it does not give the exact results in thermodynamic limit [10]. Recently by using a lengthy and complicated functional integral approach, Stasio et al. [11] have obtained the exact partition function for infinite $U$ Hubbard model for a two site systems. The two site problem can be very easily solved for the free orthofermion Hamiltonian. It is found that it gives the exact expressions for the thermodynamical quantities. This example gives one more reason in favor of the soundness of the orthofermion approach for the infinite $U$ Hubbard model.